\documentclass[aps,prd,reprint,showpacs,floatfix,nofootinbib,superscriptaddress]{revtex4-1}

\usepackage[english]{babel}
\usepackage{graphicx}
\usepackage{dcolumn}
\usepackage{amssymb}
\usepackage{amsmath}
\usepackage{amsfonts}
\usepackage{hyperref}
\usepackage{breakurl}

\begin{document}

\title{A scheme for the extraction of WIMP-nucleon scattering cross sections from total event rates}

\author{M.~Cannoni}
\affiliation{Departamento de F\'isica Aplicada, Facultad de Ciencias Experimentales, Universidad de Huelva, 21071 Huelva, Spain}
%\email{mirco.cannoni@dfa.uhu.es}

\author{J.~D.~Vergados}
\affiliation{Theoretical Physics Division, University of Ioannina, Ioannina, Gr 451 10, Greece}

\author{M.~E.~G\'omez}
\affiliation{Departamento de F\'isica Aplicada, Facultad de Ciencias Experimentales, 
Universidad de Huelva, 21071 Huelva, Spain}

\begin{abstract}

We propose a scheme that allows to analytically determine 
the three elementary cross sections and connect the solutions to the relative sign 
between the proton and the neutron spin scattering amplitudes
once the measurements of total event rate from three appropriate targets 
become available. 
In this way it is thus possible to extract the maximum information
on the supersymmetric parameter space obtainable from direct detection experiments,
in the case that the dark matter particle is the lightest neutralino.
Our scheme is based on suitably normalized form of the 
isospin momentum dependent structure functions entering in the 
spin-dependent elastic neutralino-nucleus cross section. 
We compare these functions with the commonly used ones and discuss their 
advantages: in particular, these allow in the spin-dependent cross section to factorize
the particle physics degrees of freedom from the momentum transfer dependent nuclear 
structure functions as it happens in the spin-independent cross section with the nuclear 
form factor.

\end{abstract}

	
%\date{March 17, 2011}

\maketitle

\section{Introduction}
\label{intro}

Dark matter (DM) is the dominant matter component in the Universe. 
Its nature, however, remains elusive. To settle this point is  
an issue central to both  particle physics and astrophysics.
If the dark matter halo surrounding our galaxy is composed 
of weakly interacting massive particles (WIMP), one expects
them to interact at some level with terrestrial detectors~\cite{gw},   
depositing a tiny amount of energy on the recoiling nucleus. 
From such experiments one hopes to extract the elementary WIMP-nucleon cross sections
which should provide informations on the nature of the DM particle.
Many experiments, currently running or planned, aim at achieving this goal. 

One of the most studied WIMP candidate is the lightest neutralino in the 
minimal supersymmetric standard model (MSSM) with $R$-parity conservation~\cite{JKK,BS1,BS2}.
Neutralino scattering with nucleons~\cite{or1,or2,or3,or4}
arises at the elementary level from its interactions
with quarks. 
Diagrams with the exchange of scalar particles, 
neutral CP-even Higgs bosons and squarks, give a scalar interaction 
that in the non-relativistic limit 
results into a  spin-independent (SI) interaction.
The WIMP-proton and the WIMP-neutron SI cross sections  
are almost equal, thus the scattering with the nucleus is coherent, 
i.e. the SI cross section is  proportional  to 
$A^2$,  $A$ being the atomic mass number.
Furthermore, the interaction mediated by the $Z$ boson and squarks
induces  a spin-dependent (SD) WIMP-nucleon scattering 
and a SD interaction with the whole nucleus.

Assuming that the above nucleon cross sections are of the same order, 
the SD event rate is expected to be much smaller than the SI rate, 
even in the case of fairly light nuclei, due to 
the mentioned $A^2$ dependence of the SI cross section~\cite{bedrates}.
Given that the SD cross sections carry fundamental information regarding the the underlying theory,
the natural question is, then, whether direct detection experiments can measure them.

The fact that the above discussed elementary quantities are all that can be constrained or 
extracted from direct detection experiments using three types of nuclear targets for any fixed 
WIMP mass was argued in Ref.\cite{BKKK} (see also Ref.~\cite{BK}). 

In this paper we propose a scheme that allows to analytically determine 
the three elementary cross sections and connect the solutions to the relative sign 
between the proton and the neutron spin scattering amplitudes
once the measurements of total event rate from three appropriate targets
become available, thus extracting the maximum information
on the SUSY parameter space available from direct detection experiments.

Other multi-couplings/multi-targets approaches and analysis 
are found in literature~\cite{BK}. In Ref.~\cite{bernabei} a scheme for an analysis
of the DAMA results considering both the SI-SD cross sections was proposed. In Ref.~\cite{tovey} 
the authors developed a procedure for deducing upper limits for both the spin
dependent couplings. An analysis of the experiments with SD sensitivity, in terms of both
the SD proton and neutron cross section, was performed for example
in Refs.~\cite{giuliani1,giuliani2,giuliani3,savage}. 
In some recent papers the Bayesian and frequentist statistical approach~\cite{akrami,pato} 
have been applied to the multi-target analysis to estimate the uncertainties in the 
measurements of the basic quantities. 

Our scheme is based on a suitable normalized form of the momentum dependent spin
structure functions (SSF) entering in the SD nutralino-nucleus cross section which differs 
from the standard one. Though this functions were used by one of the present authors 
in previous publications~\cite{KV,19,jdv1,jdv2,jdv3,jdv4,jdv5,jdv2010}, 
in this work, we explicitly compare the two formalisms and emphasize the advantages 
of the normalized SSF.

The plan of the paper is as follows. In Section~\ref{sec1} we discuss our scheme and 
the main results. In Section~\ref{sec2} we compare with the predictions of the 
constrained MSSM (CMSSM). 
Further discussions and comments are given in Section~\ref{sec3}. Finally, in the Appendices 
we discuss and demonstrate the advantages of the normalized spin structure functions over 
the standard formalism.

\section{An analytical scheme for extracting the elementary cross sections}
\label{sec1}

For a nucleus with mass number $A$ the total event rate
per unit of mass per unit of time is given by~\cite{jdv2010} 
\begin{equation}
R=\frac{\rho_l}{m_\chi} 
\frac{1}{Am_p}
\sqrt{\langle v^2 \rangle } \left (\sigma_{(A)}^{SI}(0) t^{SI}+ \sigma^{SD}_{(A)}(0) t^{SD} \right).
\label{rate}
\end{equation}
Here $\rho_l =0.3$ GeV/cm$^{3}$ is the local dark matter density density in the halo, 
$m_\chi$ the WIMP mass,
$m_p\approx m_n$ the nucleon  mass, i.e. $A m_p$ is the nuclear mass,
$\sigma^{SI} _{(A)}(0)$ and $\sigma^{SD} _{(A)}(0)$ are the total SI and SD 
WIMP-nucleus cross sections in the zero momentum transfer limit (ZMTL).

Assuming the near equality of the proton and neutron SI cross sections
as it is the case for neutralino, in terms of
the elementary WIMP-nucleon cross section $\sigma^{SI}$, 
the SI WIMP-nucleus cross section reads
\begin{equation}
\sigma_{({A})}^{SI}(0) = \left( \frac{\mu_A}{\mu_p} \right)^2 A^2 \sigma^{SI}, 
\label{sigmanucleusSI}
\end{equation}
with $\mu_p$ ($\mu_A$) the WIMP-proton (nucleus) reduced mass.

Moving to the SD case, the ZMTL spin nuclear matrix elements are defined as
\begin{equation}
{\Omega}^A_{p,n}  =2\sqrt{\frac{J+1}{J}} \langle \vec{S}_{p,n} \rangle,
\label{opn}
\end{equation}
with $J$  the total angular momentum of the nucleus in the ground state and
$\langle \vec{S}_{p,n} \rangle$ the expectation values of 
the spin of the proton and neutron groups.
The ZMTL SD WIMP-nucleus cross section can be written in the proton-neutron representation 
using Eqs.~(\ref{opn}),~(\ref{se0}),~(\ref{conv1}),~(\ref{stot}).
The SD WIMP-nucleon scattering amplitudes $a_p$ and $a_n$ in general
can have opposite sign (see Section~\ref{sec3}), thus
we write~\footnote
{Depending on the nucleus, also the spin matrix elements
can have opposite sign.
The relative sign thus would appear between $|{\Omega}^A_p  {a_p }|$ and 
$|{\Omega}^A_n  {a_n }|$. Anyway, the sign of ${\Omega}^A_{p,n}$ is known
from nuclear physics calculations. In particular, in case of
$^{127}$I, $^{73}$Ge, $^{19}$F used below, the relevant matrix elements are all positive.} 
\begin{equation}
\sigma_{({A})}^{SD}(0) =  \frac{\mu_A^2}{\pi}  
\left({\Omega}^A_p  {|a_p |} \pm {\Omega}^A_n {|a_n |} \right)^2 .
\label{sigmanucleusSDamp}
\end{equation}
For the scope of this paper it is convenient to rewrite Eq.~(\ref{sigmanucleusSDamp}) in 
terms of the elementary cross sections that are given by
$\sigma^{SD} _{p,n} =3 ({\mu_p}^2 /\pi) |a_{p,n}|^2$. Using this expression~\footnote
{Different normalizations for the SD cross sections
can be found in literature with
the scattering amplitudes still indicated as $a_{p,n}$. For example, 
in Ref.~\cite{BS1} one finds $\sigma^{SD} _{p,n} =12 ({\mu_p}^2 /\pi) |a_{p,n}|^2$
while in Ref.~\cite{tovey}
$\sigma^{SD} _{p,n} =24 G_F^2 ({\mu_p}^2 /\pi) |a_{p,n}|^2$;
in these cases other factors appear in Eq.~(\ref{sigmanucleusSDamp}). 
For our scope it is useful to have 
compact expressions, thus our $|a_{p,n}|^2$ include 
all numerical factors except
the factor 3, that in last analysis 
comes from the spin average.
This normalisation seems to be same 
adopted in the code \textsf{DarkSusy} used in Section~\ref{sec3} for numerical calculations. } 
and introducing $\varrho=\pm 1$, 
we get
\begin{eqnarray}
\sigma_{({A})}^{SD}(0) = \left( \frac{\mu_A}{\mu_p} \right)^2 \frac{1}{3}
\left({\Omega}^A_p  \sqrt{ \sigma^{SD}_p}
+ \varrho {{\Omega}^A_n } \sqrt{{\sigma^{SD}_n}}\right)^2.
\label{sigmanucleusSDsig}
\end{eqnarray}

The factors $t^{SI,SD}$ are the convolution of the nuclear form factor and of the momentum transfer
dependent SSF
with the WIMP velocity distribution, assumed to be the standard truncated Maxwellian 
velocity distribution. The root mean squared velocity is
$\sqrt{\langle v^2 \rangle }=\sqrt{3/2}\upsilon_0 =280$ km/s 
with $\upsilon_0 =229$ km/s the velocity
of the Sun around the center of the galaxy.
We neglect the small time dependent effects (modulation) associated with the motion of the Earth.

It is worth to remark here that
the factor $t_{SD}$ in the SD part of the total rate in Eq.~(\ref{rate})
does not depend on the particle physics couplings $a_{p,n}$ but only
on the momentum transfer dependent SSF. 
This factorization is achieved 
using the SSF $F_{ij}$ defined in Refs.~\cite{KV,19,jdv1,jdv2,jdv3,jdv4,jdv5,jdv2010} 
while it is not valid using the commonly used functions $S_{ij}$~\cite{engel,engelrev}. 
We discuss this issue  in full details in Appendix~\ref{appendixA}.

In view of the fact that the spin structure functions have been obtained in the context 
of the nuclear shell model, we 
deviate from the usual practice of dark 
matter calculations~\cite{duda}
and use also for the SI cross section
a form factor obtained in the context 
of the shell-model. Details on $t^{SI,SD}$ are given in Appendix~\ref{appendixB}.

We then proceed by first	
rewriting Eq.~(\ref{rate})
in a form where only quantities with the dimension of a cross section appear.
To this end we define the factor
\begin{eqnarray}
R_0 &=&\frac{\rho_l}{(100\text{ GeV})} \frac{1}{m_p}
\sqrt{\langle v^2 \rangle }~\sigma_0 
\simeq 1.6\times 10^{3} ~{\text{kg}}^{-1}~{\text{yr}}^{-1}\cr 
&\times& \frac{\rho_l}{(0.3 \text{ GeV/cm}^{3})}
\frac{\sqrt{\langle v^2 \rangle }}{(280 \text{ km/s})}
\frac{\sigma_0}{\text{(pb)}},
\label{R0}
\end{eqnarray}
where $\sigma_0 =10^{-8}$ pb is the current experimental 
sensitivity to the SI cross section.
$R_0$ is thus fixed to 
${R_0}\simeq 1.6\times 10^{-5} ~{\text{kg}}^{-1}~{\text{yr}}^{-1}$.
Furthermore we define the abbreviations 
\begin{eqnarray}
c^{SI} &=& \frac{(100\mbox{ GeV})}{m_\chi}\left( \frac{\mu_A}{\mu_p} \right)^2 A~t^{SI},\\
c^{SD} &=& \frac{(100\mbox{ GeV})}{m_\chi}\left( \frac{\mu_A}{\mu_p} \right)^2 \frac{1}{3} \frac{t^{SD}}{A},\\
{\mathcal R} &=& c^{SD} / c^{SI},\\
{\mathcal S} &=& \frac{R}{c^{SI} R_0}{\sigma_0}.
\label{ccoeff}
\end{eqnarray}
Leaving the dependence on the WIMP mass 
understood, Eq.~(\ref{rate}) can be rewritten in the desired form:
\begin{equation}
{\mathcal S}_A =\sigma^{SI} 
+{\mathcal R}_A \left({{\Omega}^A_p} \sqrt{ \sigma^{SD}_p}
+ \varrho {{\Omega}^A_n} \sqrt{\sigma^{SD}_n}
\right)^2 .
\label{finalrate}
\end{equation}

We now consider three nuclei with the following characteristics: 
$A_1$ for which both proton and neutron spin
matrix elements are relevant, 
a proton spin favouring target $A_{2}$ with ${\Omega}^{A_2}_p \gg {\Omega}^{A_2}_n $
and a neutron spin favouring target $A_{3}$ with
${\Omega}^{A_3}_n \gg {\Omega}^{A_3}_p $.
A system of nuclei matching these conditions is given by
three odd mass nuclei currently employed in many experiments:
$^{127}$I as the target $A_1$, $^{19}$F as the $A_{2}$ nucleus and $^{73}$Ge as the $A_{3}$.

Detailed nuclear models calculations for the spin matrix elements $\langle \vec{S}_{p,n} \rangle$ 
and form factors have been carried out in the past years.
For $^{19}$F we use the spin matrix elements of Ref.~\cite{19} that give
${\Omega}^{19}_p =1.646$ and ${\Omega}^{19}_n = -0.030$;
for $^{73}$Ge the results of Ref.~\cite{73b}, ${\Omega}^{73}_p =0.066$ and  ${\Omega}^{73}_n = 0.836$;  
finally for $^{127}$I the set obtained with the potential Bonn-A
from Ref.~\cite{127} that furnishes ${\Omega}^{127}_p =0.731$ and ${\Omega}^{127}_n =0.177$. 
The hypothesis on the  spin matrix elements are thus well satisfied.

Neglecting the neutron contribution in $A_{2}$ and the 
proton contribution in $A_{3}$,
from Eq.~(\ref{finalrate}) we have the system of equations:
\begin{eqnarray}
\left\{ \begin{array} {lcl}
\sigma^{SI} 
+{\mathcal R}_{A_1} \left({{\Omega}^{A_{1}}_p} \sqrt{ \sigma^{SD}_p} + 
\varrho {{\Omega}^{A_{1}}_n} \sqrt{\sigma^{SD}_n}
\right)^2 -{\mathcal S}_{A_1} =0\cr
 \sigma^{SI} + {\mathcal R}_{A_{2}} 
({\Omega}^{A_{2}}_p)^2 \sigma^{SD}_p 
 -{\mathcal S}_{A_{2}} =0\cr
{ \sigma_{}^{SI}} + {\mathcal R}_{A_{3}} 
({\Omega}^{A_{3}}_n)^2 \sigma^{SD}_n 
 -{\mathcal S}_{A_{3}}=0 .
\end{array} \right.
\label{system}
\end{eqnarray}
Solving for $\sigma_p^{SD}$ and $\sigma_n^{SD}$
from the second and third equations and
substituting them into the squared of the first, we obtains
a second order equation in $\sigma^{SI}$, namely
$\mathcal{A}(\sigma^{SI})^2 -2\mathcal{B}(\sigma^{SI}) +\mathcal{C}=0$.
Introducing the parameters
\begin{equation}
a ={\frac{{\mathcal R}_{A_1}}{{\mathcal R}_{A_{2}}}}
\frac{({\Omega}_p^{A_1})^2}{({\Omega}_p^{A_{2}})^2},\quad
b ={\frac{{\mathcal R}_{A_1}}{{\mathcal R}_{A_{3}}}}
\frac{({\Omega}_n^{A_1})^2} {({\Omega}_n^{A_{3}})^2}.
\label{ropar}
\end{equation}
and the abbreviations 
\begin{eqnarray}
\alpha &=& 1-a-b , \label{alpha} \\ %\;\;%\cr
\beta &=&{\mathcal S}_{A_1} - a {\mathcal S}_{A_{2}} - b  {\mathcal S}_{A_{3}} ,\\ %\;\; %\cr
\gamma &=& 2 a b ,
\label{par}
\end{eqnarray}
the coefficients are found to be
\begin{eqnarray}
\mathcal{A}&=&\alpha^2-2\gamma ,\\ %\cr
\mathcal{B}&=&\alpha \beta -\gamma\left({\mathcal S}_{A_{2}}+{\mathcal S}_{A_{3}}\right) ,\\ %\cr
\mathcal{C}&=&\beta^2 -2\gamma{\mathcal S}_{A_{2}} {\mathcal S}_{A_{3}} .
\label{Ccoeff}
\end{eqnarray}
The two sets of solutions, indicated as $s_{+}$ and $s_{-}$ respectively, are thus:
\begin{equation}
s_+ : \left\{ \begin{array} {lll}
\sigma^{SI}_{+} =\frac{\mathcal{B}+ \sqrt{{\mathcal{B}^2-\mathcal{A}\mathcal{C}}}}{\mathcal{A}}\cr
\sigma_{p,+}^{SD} =\frac{{\mathcal S}_{A_{2}}-{\sigma}_{+}^{SI}}{{\mathcal R}_{A_{2}}({\Omega}_p^{A_{2}})^2}\cr
\sigma_{n,+}^{SD} =\frac{{\mathcal S}_{A_{3}}-{\sigma}_{+}^{SI}}{{\mathcal R}_{A_{3}}({\Omega}_n^{A_{3}})^2}
\end{array} \right.
s_{-}: \left\{ \begin{array} {lcl}
\sigma^{SI}_{-} =\frac{\mathcal{B}- \sqrt{{\mathcal{B}^2-\mathcal{A}\mathcal{C}}}}{\mathcal{A}}\cr
\sigma_{p,-}^{SD} =\frac{{\mathcal S}_{A_{2}}-{\sigma}_{-}^{SI}}{{\mathcal R}_{A_{2}}({\Omega}_p^{A_{2}})^2}\cr
\sigma_{n,-}^{SD} =\frac{{\mathcal S}_{A_{3}}-{\sigma}_{-}^{SI}}{{\mathcal R}_{A_{3}}({\Omega}_n^{A_{3}})^2}.
\end{array} \right.
\label{sol}
\end{equation}

Though the solutions do not depend on $\varrho$ explicitly, for each sign the solution
is unique. In fact, each of the three equations of the system in Eq.~(\ref{system}) 
is the sum of $\sigma^{SI}$ with a term that depends on the SD cross sections,
say $\sigma^{SI}+f_i$, $i=1,2,3$. 
Obviously, for each solutions set,
the three terms $f_i$ have to be equal, that is, it must be $f_1 =f_2 =f_3$.  
However, the square in $f_1$ in the first equation contains an interference term that
can be positive or negative because depends on $\varrho$.
This means that, for example, the set $s_+$ cannot satisfy at the same time
$f_1(\varrho=+1) =f_2 =f_3$ and $f_1(\varrho=-1) =f_2 =f_3$. Given that the same argument 
applies to $s_-$,  
each set in Eq.~(\ref{sol}) is thus associated to one sign of $\varrho$.
In the specific case of $^{127}$I, $^{73}$Ge, $^{19}$F,
the factors $t^{SI,SD}$ were calculated for $E_{min}=0$ 
and neutralino masses up to 200 GeV in Ref.~\cite{jdv2010}.
Using those results we find
that  the parameters defined in Eq.~(\ref{ropar})
are very similar for all the masses,  
$a \simeq 4\times 10^{-3}$ and 
$b \simeq 1\times 10^{-2}$, that is $a\ll1$ and $b\ll1$.
Considering the cases $\mathcal{S}_1 \sim \mathcal{S}_2 \sim \mathcal{S}_3$,
$\mathcal{S}_1 \gg (\mathcal{S}_2 , \mathcal{S}_3)$ and 
$\mathcal{S}_1 \ll (\mathcal{S}_2 , \mathcal{S}_3)$,
one easily finds from Eqs.~(\ref{alpha}) and~(\ref{Ccoeff}) that
$\Delta \sim 0$, $\Delta \sim 2\gamma \mathcal{S}_1^2$, 
$\Delta \sim 2\gamma \mathcal{S}_2 \mathcal{S}_3$, respectively, thus the 
discriminant is always positive.
\begin{figure}[b!]
\includegraphics*[scale=0.45]{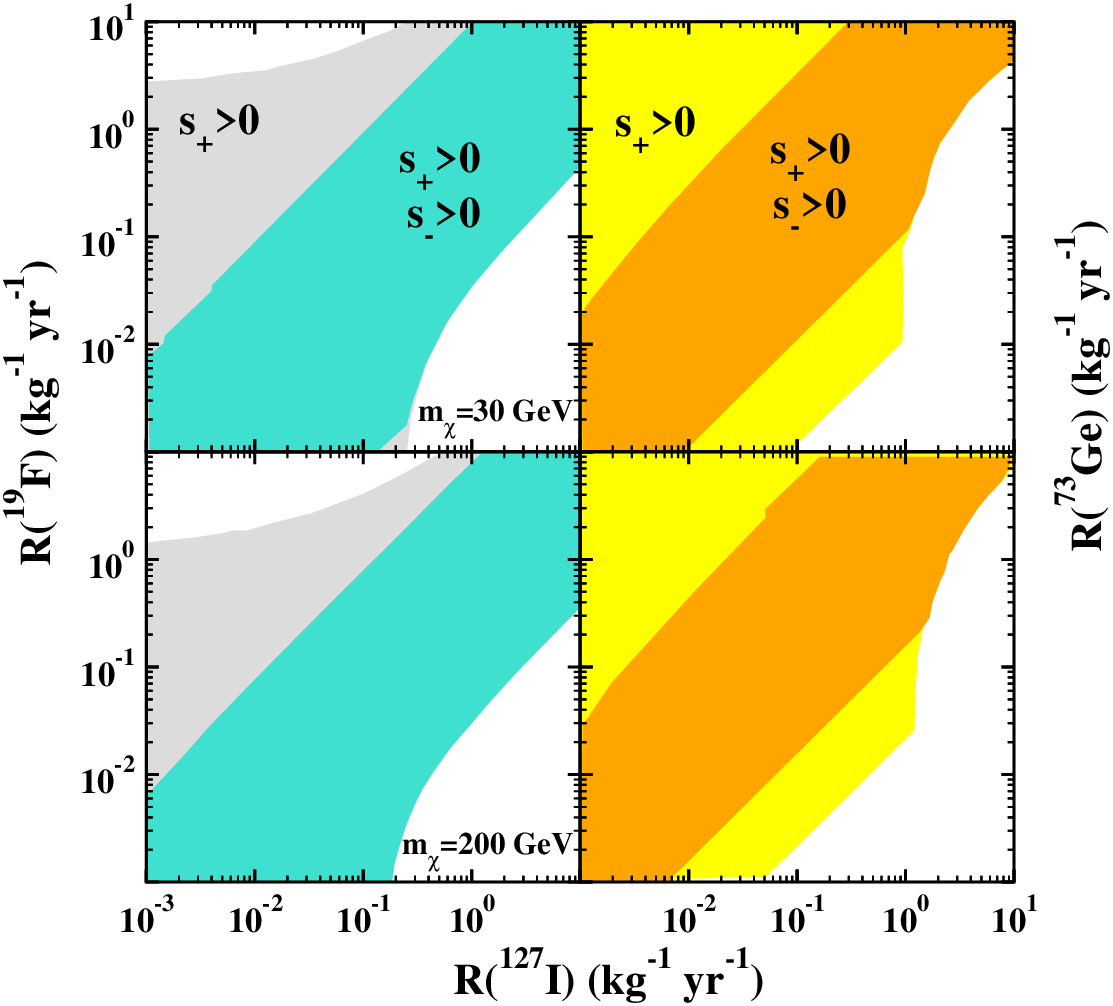}
\caption{Correlated plane $(R(^{127}\text{I}),R(^{19}\text{F}))$ (left column) and
$(R(^{127}\text{I}), R(^{73}\text{Ge}))$ (right column), 
showing the behaviour of the signs
in Eq.~(\ref{sol})
for $m_{\chi} =30$ GeV (top) and $200$ GeV (bottom).
For example, fixing a value of $R(^{127}\text{I})$ in the abscissas of the two columns, 
one can read at the same time on the ordinates, the allowed ranges 
of values for $R(^{19}\text{F})$ and $R(^{73}\text{Ge})$. 
The cross sections of $s_+$ are all positive for values of the rates 
in the coloured areas while 
the values of $s_-$ are all positive only in the restricted central turquoise and orange areas. 
In the white non-physical areas  both $s_+$ and $s_-$ have at least one negative value.}
\label{f1}
\end{figure}

Some of the cross sections in $s_+$ and $s_-$ may turn out to be negative for 
some values of the rates, thus non-physical.
We have varied numerically the three rates in four orders of magnitude
and in Fig.~\ref{f1} we show the behaviour of the signs of Eqs.~(\ref{sol}) 
in the correlated planes 
$(R(^{127}\text{I}), R(^{19}\text{F}))$, left column,
and $(R(^{127}\text{I}), R(^{73}\text{Ge}))$, right column. 
In the  central  turquoise and orange regions around the diagonal,
all the three cross sections of $s_{+}$ and all the three cross sections
of $s_{-}$ are positive.
In the grey and yellow areas, $s_{+}$ have positive values
but $s_{-}$ can have at least one negative cross section. 
In the white areas both the sets have at least one negative cross section.
As an example, suppose that a rate $R(^{127}\text{I}) =10^{-1}$ kg$^{-1}$ yr$^{-1}$ is measured
for the case $m_\chi =30$ GeV. In the interval of values 
$10^{-3} \lesssim R(^{19}\text{F}) \lesssim 6$ kg$^{-1}$ yr$^{-1}$ and 
$10^{-3} \lesssim  R(^{73}\text{Ge}) \lesssim 10$ kg$^{-1}$ yr$^{-1}$ the solutions $s_+$ are positive
but the solutions $s_-$ are all positive only in the restricted ranges
$10^{-3} \lesssim  R(^{19}\text{F})  \lesssim 8\times 10^{-1}$ kg$^{-1}$ yr$^{-1}$ and
$10^{-2} \lesssim R(^{73}\text{Ge}) \lesssim  4$ kg$^{-1}$ yr$^{-1}$. 
The variations of the allowed regions passing from $m_\chi =30$ GeV to $m_\chi =200$ GeV
are small. 
\begin{figure}[t!]
\includegraphics*[scale=0.46]{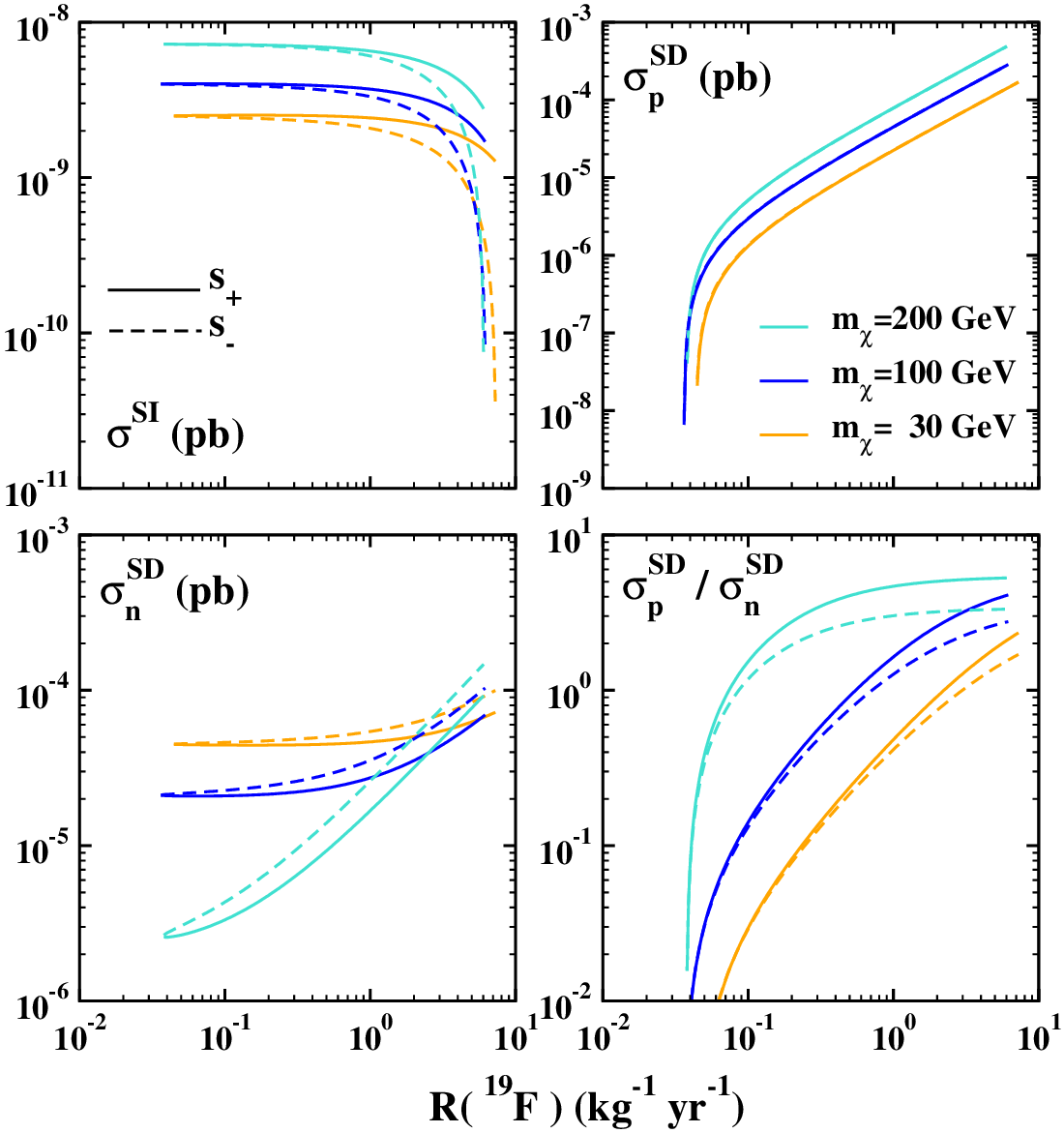}
\caption{The solutions, Eq.~(\ref{sol}), of the system Eq.~(\ref{system})
for the nuclei $^{127}$I, $^{73}$Ge, $^{19}$F as a function of $R(^{19}\text{F})$:
$s_+$, solid lines, and $s_-$, dashed lines. 
We fix $R(^{127}\text{I}) = R(^{73}\text{Ge}) =1$ kg$^{-1}$ yr$^{-1}$ and 
three values of the neutralino mass.}
\label{f2}
\end{figure}

We plot in Fig.~\ref{f2} the extracted cross sections, Eq.~(\ref{sol}),
as a function of $R(^{19}\text{F})$
fixing the rates in $^{73}\text{Ge}$ and $^{127}\text{I}$
near the experimental limits, 
$R(^{73}\text{Ge})= R(^{127}\text{I}) = 1$ kg$^{-1}$ yr$^{-1}$ and 
three values of the neutralino mass.
Note that the abscissa range
of the curves corresponds to the interval of values of $R(^{19}\text{F})$, 
where the both sets of solutions are positive, that can be read
from the left column of Fig.~\ref{f2} 
for $R(^{127}\text{I}) = 1$ kg$^{-1}$ yr$^{-1}$.
The two solutions for $\sigma^{SD}_p$ are almost overlapping:
this is because we plot the cross sections as a function of the rate in $^{19}\text{F}$
that is the nucleus more sensitive to this cross section. For the same reason
$\sigma^{SD}_p$ changes by orders of magnitude with $R(^{19}\text{F})$ while the 
variations of $\sigma^{SI}$ and $\sigma^{SD}_n$ are less dramatic. 
We find numerically that, for the chosen nuclei 
in the considered  range of rates in Fig.~\ref{f2},
$s_+$ ( $s_-$) is the set of
solutions associated to $\varrho =-1$ ($\varrho =+1$).

\section{Comparison with the CMSSM}
\label{sec2}

In the CMSSM~\cite{susylag}
all the soft breaking terms at the weak scale are obtained 
by evolving the renormalization group equations of the model
from the common scalar mass $m_0$, the common gaugino mass $m_{1/2}$, 
the common trilinear scalar coupling $A_0$, assigned at the gauge unification scale. 
The running further depends on the ratio of the vacuum expectation values of the Higgs doublet, 
$\tan\beta$, and the sign of Higgs mixing term $\mu$, which we take to be positive.

Assuming that the neutralino constitutes all the DM relic density 
as deduced from WMAP~\cite{wmap} measurements,
$0.094 < \Omega h^2 < 0.128$ at 3$\sigma$, 
the CMSSM parameter space is heavily restricted.
We further impose~\cite{gomez,cannoni} the LEP bound on the light Higgs mass $m_h >114$ GeV, 
and other phenomenological constraints
using the code \textsf{DarkSusy}~\cite{darksusy}.
The scan of the parameter space is done fixing $A_0 =0$,
$\tan\beta =10$, $50$ and varying the scalar mass and gaugino mass up to the values
$m_0<4000$ GeV, $m_{1/2} <1500$ GeV.

\begin{figure}[t!]
\includegraphics*[scale=0.46]{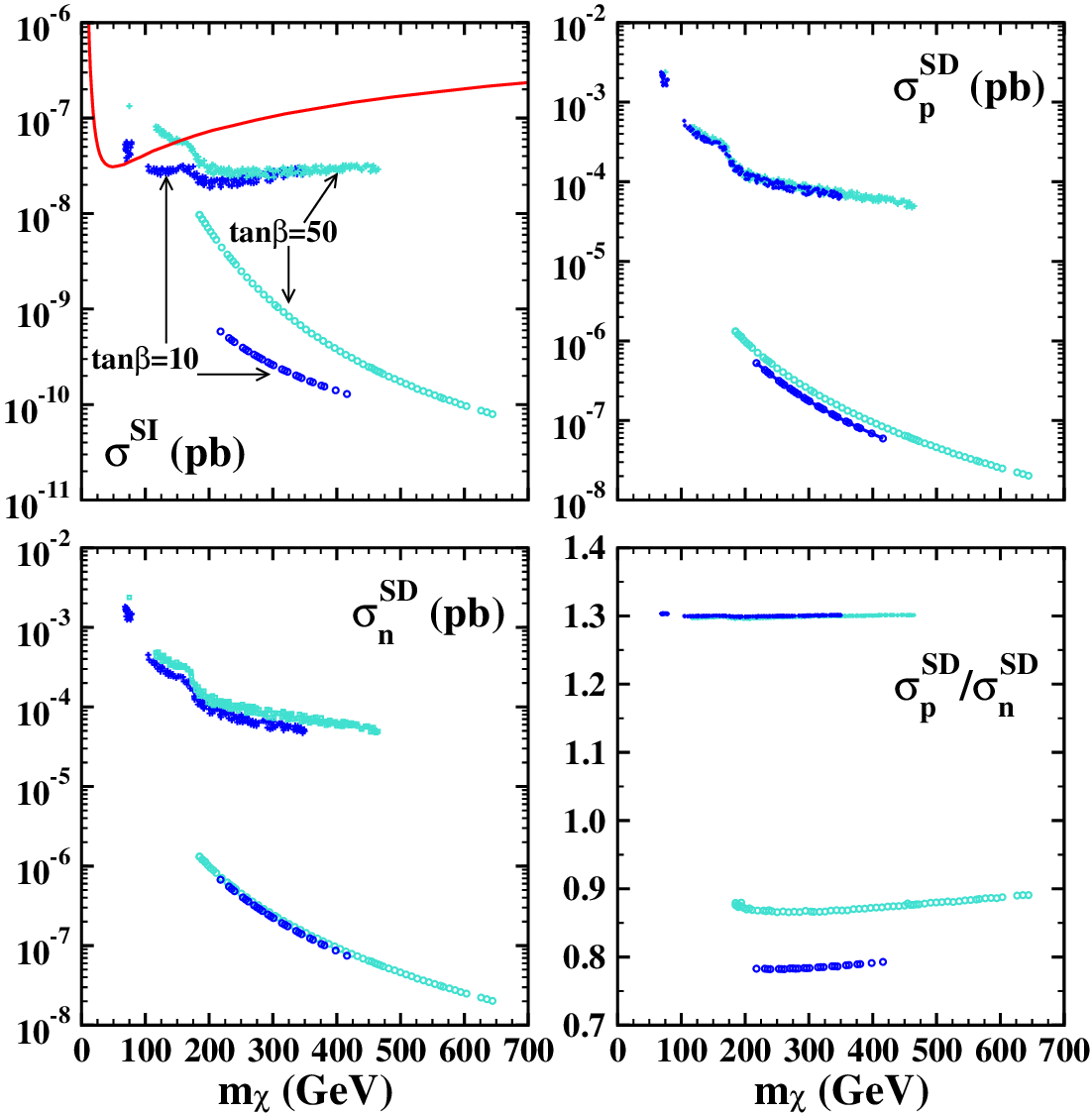}
\caption{Neutralino-nucleon cross sections as a function of the neutralino mass 
in the 
constrained minimal supersymmetric standard model 
with $A_0 =0$, $\mu>0$,  
$m_0<4000$ GeV, $m_{1/2} <1500$ GeV, $\tan\beta =10$ and 
$\tan\beta =50$. 
The points satisfy the WMAP
3$\sigma$ bound $0.094 < \Omega h^2 < 0.128$~\cite{wmap} on the relic density.
The blue  points are for $\tan\beta=10$, the turquoise ones for 
$\tan\beta=50$. The red curve is the XENON100 upper limit on the SI cross section.}
\label{f3}
\end{figure}

In Fig.~\ref{f3} we plot the SI neutralino-proton cross section (the neutron SI
cross section may differ from that for the proton by less than a few percent), 
the proton and neutron SD cross sections and 
their ratio as a function of the neutralino mass.
In all the panels the points on the bottom arise from points of the parameter space
in the stau-coannihilation region where the lightest stau has mass nearly degenerate 
with the lightest neutralino. 
The points on the top are in the focus
point regions of the parameter space, large $m_0$, where the neutralino has a relevant 
higgsino component, and some points, at large $\tan\beta$, in the funnel region,
where the neutralino mass is almost half of the mass of the neutral pseudo-scalar Higgs.
We also plot in the first panel the upper limit curve on the SI cross section obtained
by the XENON100 experiment~\cite{xenon100}, which,  at present,  has provided 
the strongest constrain on $\sigma_{SI}$. The points with the highest cross sections at 
low neutralino mass, mostly in the focus point region, are already excluded.

The proton and neutron SD scattering amplitudes, $a_p =\sum_{q} g_q \varDelta q^{(p)}$
and $a_n =\sum_{q} g_q \varDelta q^{(n)}$, are determined by    
the low energy effective couplings $g_q$ 
and $\varDelta q^{N}$, the spin fraction of the nucleon carried by the light quarks $q=u,d,s$. 
We adopted for the latter quantities the central values obtained by 
the HERMES collaboration for the proton~\cite{hermes},
$\varDelta u^{(p)} = 0.842$, $\varDelta d^{(p)} = -0.427$, $\varDelta s^{(p)} = -0.085$
and $\varDelta u^{(n)} = \varDelta d^{(p)} $, $\varDelta d^{(n)} = \varDelta u^{(p)} $,
$\varDelta s^{(n)} = \varDelta s^{(p)}$ from isospin symmetry for the neutron.
The ratio $\sigma^{SD}_p / \sigma^{SD}_n$ is almost
constant and it is greater than one in the focus point and funnel strip
and less than one in the stau-coannihilation regions.

Moreover we find that 
$a_p$ is always negative and $a_n$ always positive and this crucially depends
on the values of the spin structure functions of the proton~\cite{ellis}.
The fundamental cross sections are thus provided by the set $s_+$.
Comparing the predictions of $s_+$ in Fig.~\ref{f2}, solid lines, 
for  $m_\chi =200 $ GeV  (in the CMSSM masses below
$\sim 50$ GeV are excluded) with Fig.~\ref{f3}
we see that the extracted cross sections are compatible with 
the cross sections predicted by the CMSSM in the stau coannihilation region 
for a total rate in fluorine at the level of $5\times 10^{-2}$ kg$^{-1}$ yr$^{-1}$.

\section{Discussion}
\label{sec3}

Other realizations of the supersymmetry breaking or  
supersymmetric models other
than the CMSSM  may predict different relations 
between the elementary cross sections;
it is however clear that the SD interactions carry a lot of information on the fundamental theory.
In spite of this fact, the SD event rate can be small also in light nuclei thus rendering 
the measurement of the SD WIMP-nucleon cross sections difficult. 

We have shown that, if the total event rates are measured in 
$^{127}\text{I}$, $^{73}\text{Ge}$ and $^{19}\text{F}$ targets,
the system given by Eq.~(\ref{finalrate}) allows the extraction of all 
elementary nucleon cross sections, i.e. the spin independent as well as  
the proton and neutron  spin dependent, and that the solution are connected 
with the relative sign between the proton and neutron spin dependent 
scattering amplitudes.

These three nuclei were chosen for several reasons:
the nuclear spin matrix elements are known with sufficient degree of precision and their values
allow to write the equations for the total rates in the form of Eq.~(\ref{system}); 
the parameters defined in Eq.~(\ref{ropar})
are both $\ll 1$ and this ensures that the solutions given by Eq.~(\ref{sol}) are real for all 
values of the measured rates; finally, they are going to be employed in  
experiments with much more massive targets. 
For example, the COUPP experiment, bubble chamber based on CF$_3$I~\cite{cerdeno}, which has a proposal
for a ton scale experiment~\cite{Collar}
cannot measure the energy spectra but only counting rates (see also the discussion
in Ref.~\cite{akrami}).
This experiment contains the nuclei I and F employed in our analysis. 
If, in the future, precise rates are measured
by this experiment, in conjunction with a ton scale experiment employing Ge 
like SuperCDMS/GEODM~\cite{supercdms} and EURECA~\cite{eureca} projects, our results will
directly allow the extraction of all the cross sections, 
even if the recoil energy spectra are not available.

Certainly, due to the great variety of nuclei employed in present and planned 
experiments, other combinations of target nuclei can be used,
depending on the availability of reliable data.
If the nucleus $A_1$ has sensitivity only to the SI interaction,
it is  $a =b =0$.
From Eqs.~(\ref{alpha})-(\ref{Ccoeff}) it is found $\mathcal{A} =1$, 
$\mathcal{B} ={\mathcal S}_{A_1}$, $\mathcal{C} =\mathcal{B}^2$ thus
$\Delta=\mathcal{B}^2-\mathcal{A}\mathcal{C} =0$ and the two sets in Eq.~(\ref{sol})
reduce to the unique  solution: 
\begin{equation}
\bar{\sigma}^{SI}={{\mathcal S}_{A_1}},\;\; 
\bar{\sigma}_p^{SD} =\frac{ {\mathcal S}_{A_{2}}-{\mathcal S}_{A_1}}{{\mathcal R}_{A_{2}}({\Omega}_p^{A_{2}})^2} ,\;\;
\bar{\sigma}_n^{SD} =\frac{{\mathcal S}_{A_{3}}-{\mathcal S}_{A_1}}{{\mathcal R}_{A_{3}}({\Omega}_n^{A_{3}})^2}.
\label{degenerate}
\end{equation}
These are the expected formulas for the simpler case where
$A_1$  is an even-even nucleus, like Ar employed
in large mass experiments like WArP~\cite{WArP}, ArDM~\cite{ArDM} and 
the future project DARWIN~\cite{darwin}. 
In this case it is possible to extract the SI cross section cleanly but 
the connection of the solutions with the sign is lost. However, it can be recovered using $^{127}$I 
with other combinations like I, Ar, Ge or I, Ar, F or Xe instead of Ge.  
 
One last comment regarding the normalized SSF discussed in Appendix~\ref{appendixA} is in order.
In the usual formalism of the SD rate employing Eqs.~(\ref{dse}),~(\ref{se}), in general 
it is not possible to convert an experimental upper limit to an upper bound on $\sigma^{SD}_{p,n}$  
independently of neutralino properties~\cite{bottino} because the nuclear momentum dependent
degrees of freedom are not decoupled from the particle physics couplings.
As a matter of fact, has become a standard procedure to use the method of Ref.~\cite{tovey}
to set limits on the cross sections in a WIMP-independent way. This independence is anyway 
achieved using the ZMTL total cross section in Eq.~(\ref{stot}) and not the full $q$-dependent
cross section: this can be a bad approximation in the case of heavy nuclei and/or heavy
neutralinos. The correct $q$-dependence, at least for the nuclei for which full nuclear physics 
calculations are available, can be incorporated in this method employing the 
normalized SSF which allow the factorization of the particle physics degrees of freedom 
from the $q$-dependent nuclear physics structure functions.

\acknowledgements
M.~C. is a MultiDark fellow. 
The authors acknowledge support from the 
Spanish MICINN projects Consolider-Ingenio 2010 
Programme, grant MultiDark CSD2009-00064 and FPA2009-10773, and  
from Junta de Andalucia under grant P07FQM02962.

\appendix
\section{The normalized spin structure functions}
\label{appendixA}

The calculation of the elastic neutralino-nucleus full momentum transfer dependent 
cross section is conveniently performed in the isospin basis.
In the formalism introduced by in Ref.~\cite{engel} and reviewed in~\cite{engelrev},
which is the commonly used in the particle physics and astrophysics literature 
(see the reviews~\cite{JKK,BS1,BS2}), 
the differential cross section is written as
\begin{equation}
\frac{d\sigma^{SD}_{(A)}}{dq^2} =\frac{1}{4(\mu_A v)^2} \sigma^{SD}_{(A)}(0) \frac{S(q)}{S(0)}.
\label{dse}
\end{equation}
The momentum transfer dependent SSF  enter in the factor
\begin{equation}
S(q) = a^2_0 S_{00}(q) +  a_0 a_1 S_{01}(q) + a^2_1 S_{11}(q) .
\label{se}
\end{equation}
In the ZMTL reduces to
\begin{eqnarray}
&S(0)& = \frac{2J+1}{\pi}J(J+1) \cr
&\times& \left[\frac{ a_0  (\langle \vec{S}_p \rangle + \langle \vec{S}_n \rangle )
 + a_1 (\langle \vec{S}_p \rangle - \langle \vec{S}_n \rangle )}{2 J}\right]^2 ,
\label{se0}
\end{eqnarray}
being $a_{0,1}$ the isoscalar and isovector coupling determined by the 
particle physics model and related to the proton-neutron representation 
by the relations
\begin{equation}
a_0 = a_p + a_n,\;\;\;\; a_1 = a_p - a_n . 
\label{conv1}
\end{equation}
In the ZMTL the total cross section is
\begin{equation}
\sigma^{SD}_{(A)}(0) = \frac{4\mu^2_A}{2J+1} S(0) .
\label{stot}
\end{equation}
The single functions $S_{ij}(q)$, $i,j=0,1$ denoting the isospin channels,
that are obtained from nuclear shell-model calculations,
are not normalised to one, but the normalisation of the ''spin form factor'' 
is achieved by the ratio $S(q)/S(0)$.

In the formalism of Refs.~\cite{KV,19},
the SSF are given by:
\begin{equation}
F_{ij}(u) = \sum_{\lambda,k} 
\frac{\Omega^{(\lambda,k)}_i (u) \Omega^{(\lambda,k)}_j (u)} {\Omega_i \Omega_j},
\label{Fij}
\end{equation}
where $\Omega^{(\lambda,k)}_i (u)$ are the matrix elements 
of the multi-pole expansion of the nuclear spin operators 
evaluated in a shell-model multiparticle basis and 
$\Omega^{(0,1)}_i (0)\equiv \Omega_i$ are the ZMTL values.
Details can be found in Ref.~\cite{19}.

The dimensionless variable $u$ is given by $u=q^2 b^2 /2$ where $b$
is the nuclear oscillator size parameter that depends on the atomic mass number
of the particular nucleus, 
$b=1\,\text{fm}\,A^{1/6} \approx 5\times 10^{-3} \,\text{MeV}^{-1} \,A^{1/6}$.
The differential neutralino-nucleus cross section in the laboratory frame reads 
\begin{equation}
\frac{d\sigma^{SD}_{(A)}}{du} =\frac{1}{2(\mu_A b v)^2} \frac{4\mu^2_A	}{2J+1} \mathcal{F}(u),
\label{dsv}
\end{equation}
with
\begin{eqnarray}
\mathcal{F}(u)= \frac{2J+1}{16\pi} 
&&(a^2_0 \Omega^2_0 F_{00}(u)
+ 2 a_0 a_1 \Omega_0 \Omega_1 F_{01}(u) \cr
&&+ a^2_1 \Omega^2_1 F_{11}(u) ).
\label{sv}
\end{eqnarray}
The SSF defined Eq.~(\ref{Fij}) are normalized to one at $u=0$ because the  
ZMTL values $\Omega_{0,1}$ has been factored out thus explicitly appearing in Eq.~(\ref{sv}).
These are related to the $\Omega_{p,n}$ defined in Eq.~(\ref{opn}) by
\begin{equation}
\Omega_0 =\Omega_p + \Omega_n,\;\; \Omega_1 =\Omega_p - \Omega_n .
\label{conv2}
\end{equation}

The formulas connecting the two sets of SSF are:
\begin{eqnarray}
S_{00}(q)&=& \frac{2J+1}{16\pi} \Omega^2_0 F_{00}(u), 
\label{s00}\\
S_{01}(q)&=& \frac{2J+1}{8\pi} \Omega_0 \Omega_1 F_{01}(u),\\
S_{11}(q)&=& \frac{2J+1}{16\pi} \Omega^2_1 F_{11}(u).
\label{ssf}
\end{eqnarray}

\begin{figure}[t!]
\includegraphics*[scale=0.44]{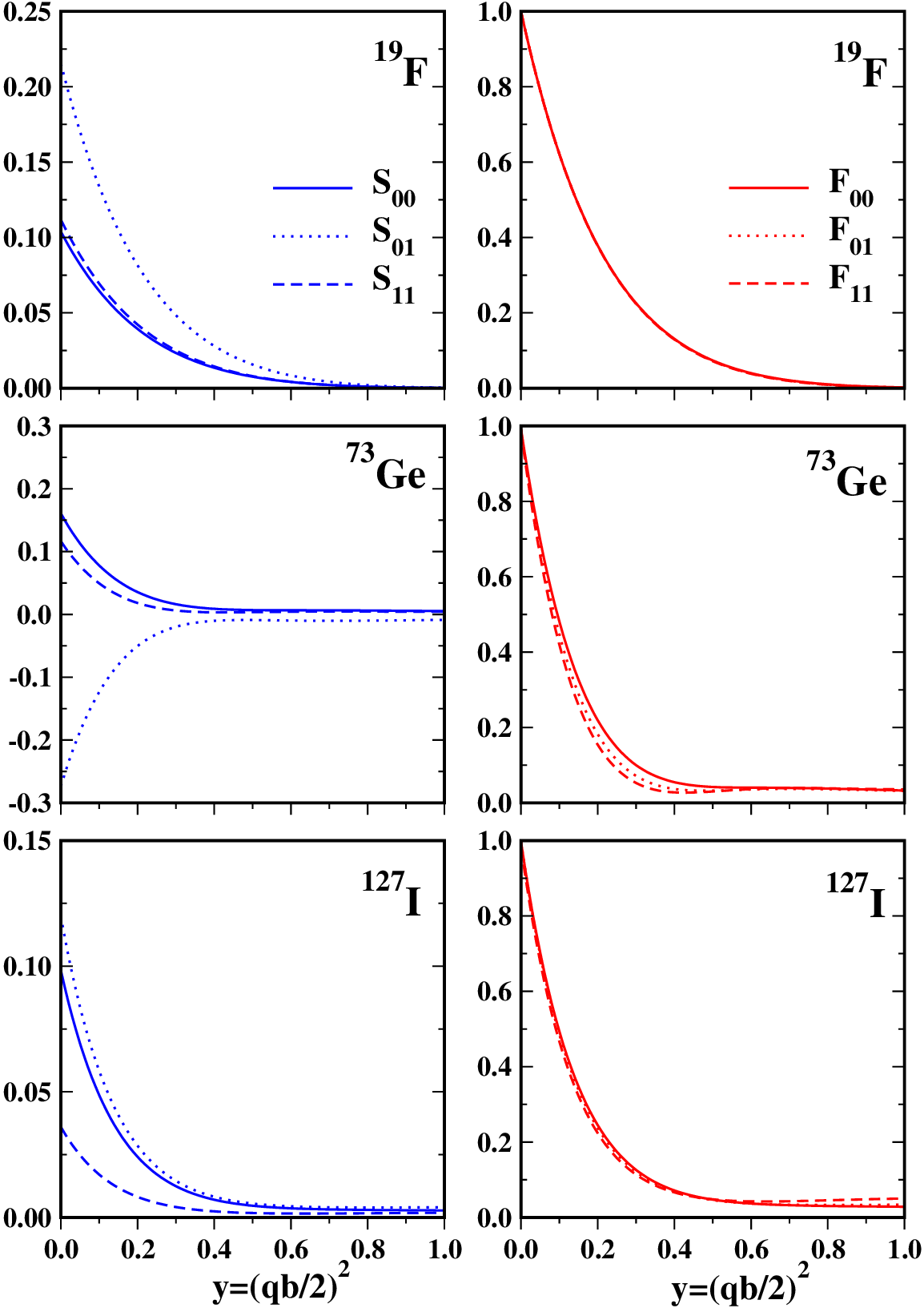}
\caption{The momentum transfer dependent spin structure functions $S_{ij}$ (left column) and $F_{ij}$ (right column),
from top to bottom for the nuclei $^{19}$F, $^{73}$Ge, $^{127}$I.}
\label{figSSF}
\end{figure}

We are in the position to appreciate the differences between the two formalisms.
In the case of the light nucleus $^{19}$F, the functions $F_{ij}$ are given
in an analytical form in Refs.~\cite{19,jdv2010} as
\begin{equation}
F^{(^{19}\text{F})}_{ij}(u) = e^{-u}\sum\limits_{k=0}^{4} {\textsf{f}}^{(k)}_{ij} u^k .
\label{p19}
\end{equation}  
For the isotope $^{73}$Ge, the authors of Ref.~\cite{73b} furnish a polynomial fit 
to the calculated SSF of the following form: 
\begin{equation}
S^{(^{73}\text{Ge})}_{ij}(y)= \sum\limits_{k=0}^{6} \textsf{g}^{(k)}_{ij} y^k ,
\label{p73}
\end{equation}
while for $^{127}$I, the authors of Ref.~\cite{127} fit their functions 
(here we use the ones obtained with the Bonn-A nucleon-nucleon potential) with 
\begin{equation}
S^{(^{127}\text{I})}_{ij} (y) = e^{-2y} \sum\limits_{k=0}^{8} \textsf{i}^{(k)}_{ij} y^k ,
\label{p127}
\end{equation}
in terms of the variable $y=(qb/2)^2 =u/2$.
The coefficients of the polynomials are found in the cited papers 
and in the review~\cite{BS2}. Using Eqs.~(\ref{s00})-(\ref{p127})
and the ZMTL $\Omega_{0,1}$ obtained with Eqs.~(\ref{opn}) and~(\ref{conv2}),
we plot in Figure~\ref{figSSF} the two sets for the nuclei 
$^{19}$F, $^{73}$Ge, $^{127}$I as a function of the variable $y$. 
Introducing the quantity $Q_0  =(b^2 A m_p)^{-1}$ the variables $u$ and $y$
are related to the recoil energy through the relation
\begin{equation}
E_R = u Q_0 =2y Q_0 =2y \times 40\,A^{-4/3}\text{MeV}.
\label{erecoil}
\end{equation} 

As it is evident, while the $S_{ij}$ look quite different, the $F_{ij}$ are practically 
identical in the interesting range $0<y<1$: using Eq.~(\ref{erecoil}), $y=1$ corresponds
to $E_R =1578$ keV in $^{19}$F, $E_R =262$ keV in $^{73}$Ge and $E_R =125$ keV in $^{127}$I.
We remark that while for the light nucleus $^{19}$F the functions in Eq.~(\ref{p19})
are an exact result of the shell-model calculations, the functions in Eqs.~(\ref{p73}),~(\ref{p127})
for the more complex nuclei $^{73}$Ge, $^{127}$I, are fits to the results of the shell-model
calculations which cannot be cast in a simple analytical form.
Furthermore, in a more recent 
large-scale shell-model computation performed in Ref.~\cite{toivanen} 
the authors calculate the $F_{ij}(u)$ SSF for the heavy nuclei 
$^{127}$I, $^{129}$Xe, $^{131}$Xe, $^{133}$Cs and find the same behaviour
(see Fig.~3 and Fig.~4 of~\cite{toivanen}).  

Given that $F_{00}(u)\simeq F_{01}(u) \simeq F_{11}(u)$ we can describe the momentum dependence 
of the cross section in terms of {\it only one} structure function, for example the one 
associated with the isovector channel $F_{11}(u)$. 
Considering the intrinsic uncertainties in the nuclear physics calculations, 
in the WIMP velocity distribution function and in the particle physics couplings, 
this is by far a well motivated approximation with negligible error in the calculation
of the cross section. 
We thus rewrite Eq.~(\ref{sv}) as
\begin{equation}
\mathcal{F}(u)= \frac{2J+1}{16\pi}
( a_0 \Omega_0 + a_1 \Omega_1)^2 F_{11}(u),
\end{equation}
and going to the proton-neutron representation using 
Eqs.~(\ref{se0}),~(\ref{conv1}),~(\ref{stot}),~(\ref{conv2}), we  finally find
\begin{equation}
\frac{d\sigma^{SD}_{(A)}}{du} =\frac{1}{2(\mu_A bv)^2} \sigma^{SD}_{(A)} (0)F_{11}(u).
\label{dsv1}
\end{equation}
Eq.~(\ref{dsv1}) is equivalent to, but simpler than, Eq.~(\ref{dse}).
We summarize and highlight the advantages of the normalized SSF:
\begin{enumerate}
\item 
the three $F_{ij}$ are normalized to one at zero momentum transfer and are 
always positive definite, while the interference term $S_{01}$, 
depending on the nucleus, 
can be negative (see the case 
of $^{73}$Ge);
\item 
the three $F_{ij}$ are to a very good approximation identical, thus one has to deal 
only with one SSF instead of three;
\item
in Eq.~(\ref{dsv}), differently from Eq.~(\ref{dse}), the particle physics and nuclear 
physics momentum dependent degrees of freedom are thus completely factorized.
\end{enumerate}

\section{The factors $t^{SI,SD}$}
\label{appendixB}

Starting from Eq.~(\ref{dsv}) 
the $t^{SD}$ factor in Eq.~(\ref{rate}) is defined as
\begin{equation}
t^{SD}=\int\limits_{u_{min}}^{u_{max}} du \,F_{11}(u) \int\limits_{v_{min}(u)}^{v_{max}} d^{3} \vec{v} 
\frac{v}{\sqrt{\langle v^2 \rangle }}\frac{f(\vec{v})}{2(\mu_A bv)^2}.
\label{tsd}
\end{equation}

As stated in Section~\ref{sec1}, we neglect in this work the small effects associated 
to motion of the Earth and consider the standard WIMP Maxwellian velocity distribution 
truncated at the escape velocity $v_{esc}$~\cite{lewinsmith}. 
The integration limits are: $v_{max} =v_{esc}$,
$v_{min}(u) =\sqrt{(u/(2\mu^2_A b^2)}$, $u_{max} =2 (\mu_A b v_{esc})^2$ and $u_{min}={E_{min}/Q_0}$,
where $E_{min}$ is set by the energy threshold of the detector.

The factor $t^{SI}$ is given by the same formula replacing $F_{11}(u)$ with 
$|F(u)|^2$, where $F{(u)}$ is the nuclear form factor~\cite{KSnuc,KV,19}.
For consistency we use the same 
nuclear model, i.e. the same orbitals, the same interaction and the 
same size parameter, in obtaining the spin structure functions and the 
coherent form factor. In the context of the shell-model, the coherent form factor arises:
(1) from the contribution of the $A_{\mbox{\tiny{core}}} = A-A_{\mbox{\tiny{exc}}}$ 
nucleons in the core. These nucleons are put in the lowest orbitals allowed by the Pauli 
principle;
(2) from the contribution of the nucleons $A_{\mbox{\tiny{exc}}}$ outside of the closed 
shell. The wave function describing them is  obtained by diagonalizing a suitable effective 
interaction in the corresponding model space.
It turns out that this last contribution can be very well approximated by allowing 
the  $A_{\mbox{\tiny{exc}}}$ nucleons to be placed in the lowest energy harmonic 
oscillator shells allowed by the Pauli principle. This approximation is quite good 
even in the case of light nuclei~\cite{19}.

\end{document}